\documentclass[letterpaper, 10 pt, conference]{ieeeconf}
\IEEEoverridecommandlockouts
\newcommand{\comment}[1]{}

\newcommand{\be}[0]{\begin{equation}}
\newcommand{\ee}[0]{\end{equation}}
\newcommand{\ben}[0]{\begin{equation*}}
\newcommand{\een}[0]{\end{equation*}}
\newcommand{\bena}[0]{\begin{eqnarray*}}
\newcommand{\eena}[0]{\end{eqnarray*}}
\newcommand{\bea}[0]{\begin{eqnarray}}
\newcommand{\eea}[0]{\end{eqnarray}}

\usepackage[english]{babel}
\usepackage{amsthm}
\usepackage[version=4]{mhchem}
\usepackage{siunitx}
\usepackage{array}
\usepackage{longtable}%tabulary
\usepackage{nomencl}
\usepackage{mathrsfs} 
\usepackage{mathtools} 
\usepackage{optidef}
\usepackage{multicol}
\usepackage{svg}
\usepackage{lipsum}% http://ctan.org/pkg/lipsum
\usepackage{calrsfs}
\usepackage{scalerel}
\usepackage{stackengine}
\usepackage{booktabs}
\usepackage{caption}
\usepackage{enumerate}
\usepackage{cite}
\usepackage{amssymb,amsfonts}
\usepackage{algorithmic}
\usepackage{graphicx}
\usepackage{textcomp}

\DeclareMathAlphabet{\mathcal}{OMS}{cmsy}{m}{n}
\theoremstyle{definition}
\newtheorem{definition}{Definition}
\newtheorem{theorem}{Theorem}
\begin{document}

\title{Decentralized CBF-based Safety Filters for Collision Avoidance of Cooperative Missile Systems with Input Constraints}
\author{Johannes Autenrieb$^{1}$, Mark Spiller$^{1}$
\thanks{$^{1}$ German Aerospace Center (DLR), Institute of Flight Systems, 38108, Braunschweig, Germany.
(email: \texttt{johannes.autenrieb@dlr.de, mark.spiller@dlr.de})}
}

\maketitle

\begin{abstract}
This paper presents a decentralized safety filter for collision avoidance in multi-agent aerospace interception scenarios. The approach leverages robust control barrier functions (RCBFs) to guarantee forward invariance of safe sets under bounded inputs and high-relative-degree dynamics. Each effector executes its nominal cooperative guidance command, while a local quadratic program (QP) modifies the input only when necessary. Event-triggered activation based on range and zero-effort miss (ZEM) criteria ensures scalability by restricting active constraints to relevant neighbors. To ensure feasibility under multiple simultaneously active constraints, a slack-variable relaxation scheme is introduced that prioritizes critical agents in a Pareto-optimal manner. Simulation results in many-on-many interception scenarios demonstrate that the proposed framework maintains collision-free operation with minimal deviation from nominal guidance, providing a computationally efficient and scalable solution for safety-critical multi-agent aerospace systems.

\end{abstract}

%===============================================================================
\section{Introduction}
Modern aerospace systems increasingly rely on coordinated autonomous operations to accomplish complex missions under strict temporal and spatial constraints. Cooperative engagement scenarios—where multiple effectors must interact with targets while maintaining safe separation—pose significant challenges for real-time control synthesis. Applications include formation flying, distributed sensing, and time-critical interception missions, where rapid maneuvers in close proximity are unavoidable. These scenarios demand control strategies that guarantee collision-free operation while preserving mission effectiveness under bounded actuation and limited sensing.

\begin{figure}
    \centering
    \includegraphics[width=\linewidth]{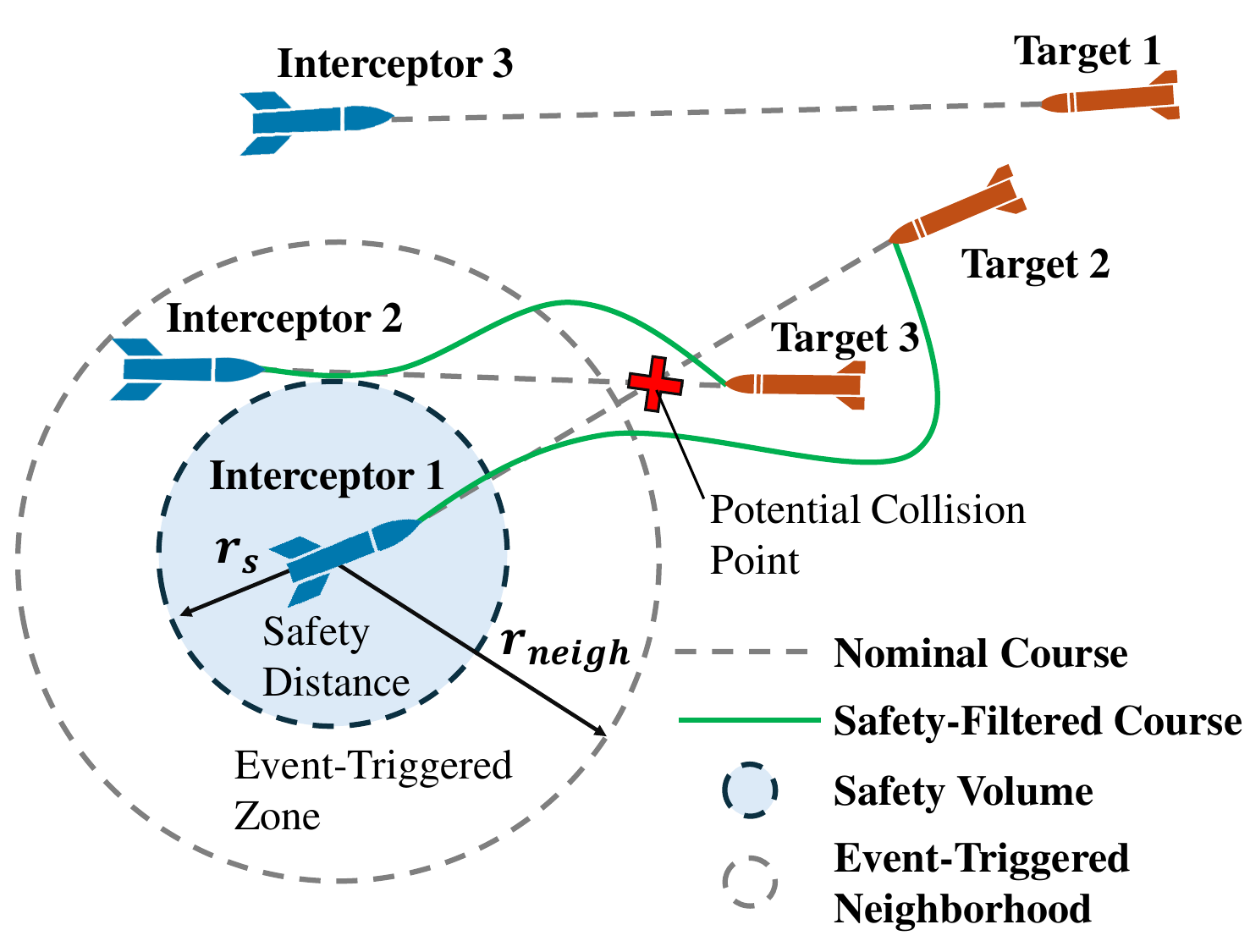}
    \caption{Illustration of decentralized, event-triggered safety filtering in a many-on-many interception mission with cooperative effectors.}
    \label{fig:placeholder}
\end{figure}

In cooperative missile guidance, these challenges are particularly acute. Multiple interceptors must engage one or more targets, often with coordinated timing requirements to maximize success \cite{LI2024109212,Li2023}. At the engagement level, cooperative guidance schemes typically combine high-level target assignment with local trajectory generation. While effective for distributed decision-making, such approaches often neglect detailed interactions between neighboring effectors. As a result, even optimally coordinated assignments may yield conflicting trajectories, where interceptors converge and risk mid-air collisions despite pursuing distinct targets \cite{Lv2024}. The central difficulty lies in bridging the gap between discrete assignment decisions at the mission-planning level and continuous-time collision avoidance during execution.

This challenge extends to safety-critical multi-agent systems more broadly, where decentralized agents must coordinate under limited communication and sensing while respecting hard safety constraints. Control Barrier Functions (CBFs) provide a principled framework for enforcing safety by certifying forward invariance of safe sets \cite{ames2016control}. Recent work has applied barrier-function methods to real-time control, enabling constraint-satisfying modifications of nominal inputs \cite{BORRMANN2015UAV}. However, most formulations assume geometric braking or idealized conditions and do not systematically address bounded inputs or feasibility issues arising in simultaneous multi-agent interactions.

A key difficulty is the ``leaky corner problem,'' which occurs when pairwise barrier sets intersect such that no admissible control input satisfies all constraints. Prior work has shown that the intersection of individual pairwise safe sets does not generally represent the true joint safe set, and identifying such infeasible regions without full reachability analysis remains an open problem \cite{Aloor_2025}. Despite recognition of this issue, existing approaches either rely on computationally expensive global methods or neglect it entirely, limiting applicability to real-time aerospace systems.

This paper proposes a decentralized, event-triggered safety filter for cooperative guidance based on robust control barrier functions (RCBFs) \cite{breeden2023robust}. The formulation explicitly incorporates input constraints in a high-relative-degree setting and ensures feasible safety decisions even with multiple conflicting agents. Each effector solves a local quadratic program (QP) that modifies its nominal guidance command only when necessary, using information from a bounded neighborhood defined by range and zero-effort miss (ZEM) criteria. To handle feasibility challenges from simultaneous constraints, we introduce a slack-variable prioritization mechanism that allocates relaxation selectively to less critical neighbors, ensuring imminent threats remain strictly enforced (see Fig.~\ref{fig:placeholder}).

%The contributions of this paper are threefold: First, we propose a decentralized event-triggered safety filter for cooperative aerospace interception missions based on robust high-order barrier functions that explicitly account for input saturation and provide formal feasibility conditions for the one-versus-one collision avoidance case. Second, we introduce a slack-variable relaxation scheme with prioritized constraint enforcement that ensures feasibility under conflicting multi-agent constraints and provides Pareto-optimal resolution of competing safety objectives. Third, we develop an event-triggered activation strategy using range and ZEM gates that reduces computational demand while preserving safety guarantees, enabling scalability to large multi-agent systems.

The proposed architecture enforces provable safety while preserving the performance of the nominal cooperative guidance scheme to the maximum extent possible. It provides a systematic and computationally efficient solution to decentralized collision avoidance in multi-agent aerospace engagements, with applicability to a broad class of aerospace and robotics scenarios where safety under input constraints is critical.
\section{Preliminaries}
Consider the input-affine nonlinear system
\begin{align}
	\dot{\mathbf{x}}
	=
	f(\mathbf{x}) + g(\mathbf{x})\mathbf{u} + g(\mathbf{x})\mathbf{w},
    \label{NonlinearPlant1}
\end{align}
with states \(\mathbf{x} \in \mathbb{R}^n\), constrained control inputs \(\mathbf{u}\in \mathcal{U}\subset \mathbb{R}^m\), and input disturbances \(\mathbf{w}\in \mathcal{W}\subset \mathbb{R}^m\), where \(\mathcal{U}\) and \(\mathcal{W}\) are compact. The functions \(f:\mathbb{R}^n \to \mathbb{R}^n\) and \(g:\mathbb{R}^n \to \mathbb{R}^{n\times m}\) are assumed to be locally Lipschitz continuous in \(\mathbf{x}\). 

The states are constrained by a function \(h:[t_0,\infty)\times \mathbb{R}^n \to \mathbb{R}\) with \(h\in \mathcal{G}^2\), where \(\mathcal{G}^2\) denotes the set of functions that are twice differentiable in time and of relative degree two, in the sense that the inputs \(\mathbf{u}\) and \(\mathbf{w}\) appear explicitly in the second derivative of \(h\). Based on \(h\), the time-varying zero-sublevel set is defined as
\begin{align}
	\mathcal{S}(t)=\{\mathbf{x} \in \mathbb{R}^n \mid h(t,\mathbf{x}) \le 0\}.
\end{align}

The following standard definitions are recalled \cite{Nagumo_1942, Blanchini_1999}:
\begin{definition}
A set \(\mathcal{S} \subset \mathbb{R}^n\) is \emph{positively invariant} for the system \eqref{NonlinearPlant1} if, for every \(\mathbf{x}_0 \in \mathcal{S}\), the solution satisfies \(\mathbf{x}(t) \in \mathcal{S}\) for \(\mathbf{x}(0) = \mathbf{x}_0\) and all \(t \in I(\mathbf{x}_0) = [0,\tau_{\max} = \infty)\).
\end{definition}
\begin{definition}
A continuous function \(\alpha: (-b, a) \rightarrow \mathbb{R}\), with \(a,b > 0\), is an \emph{extended class-\(\mathcal{K}\) function} (\(\alpha \in \mathcal{K}\)) if \(\alpha(0) = 0\) and \(\alpha\) is strictly monotonically increasing. If \(a = b = \infty\), \(\lim_{r \rightarrow \infty} \alpha(r) = \infty\), and \(\lim_{r \rightarrow -\infty} \alpha(r) = -\infty\), then \(\alpha\) is a \emph{class-\(\mathcal{K}_\infty\) function} (\(\alpha \in \mathcal{K}_\infty\)).
\end{definition}

The objective is to guarantee forward invariance of a subset of \(\mathcal{S}(t)\) in order to satisfy the constraints described by \(h\), despite input limitations and disturbances. This problem is studied in detail in \cite{breeden2023robust}, where a solution is provided through \emph{robust control barrier functions (RCBFs)}. The following definition is adapted from Theorem~9 in \cite{breeden2023robust}.

% \textcolor{blue}{Comment: I think the choice on  $\xi$ is not the best one, because it is heavily used in the CBF community in a differnt context. Do you think it would be fine to use a different variable, such as $\xi$, $\xi$ or $\zeta$?} 

\begin{theorem}[{\cite[Theorem~9]{breeden2023robust}}]
Suppose \(h\in\mathcal{G}^2\) and there exists {\(\xi>0\)} such that for all \((t,\mathbf{x})\in [t_0,\infty)\times \mathcal{S}(t)\),
\begin{align}
	\max_{\mathbf{w}\in \mathcal{W}}
	\left(
	\inf_{\mathbf{u}\in \mathcal{U}}
	\ddot h(t,\mathbf{x},\mathbf{u},\mathbf{w})
	\right)
	\le -{\xi}.
    \label{num:rcbf_max_acc}
\end{align}
Then the function
\begin{align}
	H(t,\mathbf{x}) = h(t,\mathbf{x}) + \frac{|\dot h(t,\mathbf{x})|\dot h(t,\mathbf{x})}{2{\xi}}
\end{align}
is a RCBF on the set
\begin{align}
	\mathcal{S}^{res}_H(t)
	=
	\{\mathbf{x} \in \mathbb{R}^n \mid 
	H(t,\mathbf{x})\le 0
	\land
	h(t,\mathbf{x})\le 0
	\}
	\subseteq
	\mathcal{S}(t)\notag,
\end{align}
i.e.,
\begin{align}
	\max_{\mathbf{w}\in \mathcal{W}}
	\left[
	\inf_{\mathbf{u}\in \mathcal{U}}
	\dot H(t,\mathbf{x},\mathbf{u},\mathbf{w})
	\right]
	\le
	\alpha (-H(t,\mathbf{x}))
\end{align}
for any class-\(\mathcal{K}\) function \(\alpha\) and all \(t\in[t_0,\infty)\), \(\mathbf{x}\in\mathcal{S}^{res}_H(t)\). The set
\begin{align}
	U(t,\mathbf{x})=
	\Bigl\{
	\mathbf{u}\in\mathcal{U}
	\;\big|\;
	\max_{\mathbf{w}\in \mathcal{W}}
	\dot H(t,\mathbf{x},\mathbf{u},\mathbf{w})
	\le
	\alpha (-H(t,\mathbf{x}))
	\Bigr\}
	\notag
\end{align}
is non-empty for any class-\(\mathcal{K}\) function \(\alpha\) and any \((t,\mathbf{x})\in[t_0,\infty)\times\mathcal{S}^{res}_H(t)\). Moreover, any control law \(\mathbf{u}(t,\mathbf{x})\) that is locally Lipschitz continuous in \(\mathbf{x}\) and piecewise continuous in \(t\), and satisfies \(\mathbf{u}(t,\mathbf{x}) \in U(t,\mathbf{x})\), renders \(\mathcal{S}^{res}_H\) forward invariant.
\label{num:theorem_theo9_rcbf_paper}
\end{theorem}

\section{Problem Formulation}
Let \( \mathcal{M} = \{1, 2, \dots, N\} \) denote the index set of \( N \) effectors. 
For simplicity, the dynamics of each agent \( i \in \mathcal{M} \) are described by a second-order integrator model in three-dimensional Euclidean space. Specifically, the state vector \( \mathbf{x}_i \in \mathbb{R}^6 \) evolves according to the following linear time-invariant system:
\begin{equation}
    \dot{\mathbf{x}}_i = 
    \begin{bmatrix}
        \dot{\mathbf{p}}_i \\ \dot{\mathbf{v}}_i
    \end{bmatrix}
    =
    \begin{bmatrix}
        0 & \mathbf{I} \\
        0 & 0
    \end{bmatrix}
    \begin{bmatrix}
        \mathbf{p}_i \\ \mathbf{v}_i
    \end{bmatrix}
    +
    \begin{bmatrix}
        0 \\ \mathbf{I}
    \end{bmatrix}
    {\mathbf{a}_i},
\end{equation}
where \( \mathbf{p}_i \in \mathbb{R}^3 \) denotes the position of agent \( i \), \( \mathbf{v}_i \in \mathbb{R}^3 \) its velocity, and \( {\mathbf{a}_i} \in \mathbb{R}^3 \) the control input representing the commanded acceleration vector.

The agents are modeled as point masses subject to control and physical constraints, including input saturation. We assume the following physical bounds:
\begin{equation}
    \|\mathbf{v}_i\| \leq v_{\max}, 
    \qquad 
    \|{\mathbf{a}_i}\| \leq {a_{\max}}.
\end{equation}

Define the relative quantities between two agents \( i \) and \( j \) as
\begin{equation}
    \mathbf{r}_{ij} = \mathbf{p}_i - \mathbf{p}_j, 
    \qquad 
    \mathbf{v}_{ij} = \mathbf{v}_i - \mathbf{v}_j.
\end{equation}

The mission objective is target interception in a many-on-many scenario, where multiple incoming threats must be countered by multiple defending effectors. Each effector is assigned to a target through a cooperative Weapon-Target Assignment (WTA) system. While dynamic WTA schemes can recompute assignments in real time to adapt to scenario changes, practical implementations rely on efficient iterative or linearized solvers to meet the stringent real-time constraints of embedded systems \cite{autenrieb2025WTA}. These methods optimize assignments effectively but introduce simplifications that relax optimality guarantees and often neglect inter-effector interactions during the engagement \cite{Ostermann2026WTA}. As a result, even combinatorially optimal assignments may generate conflicting trajectories, risking mid-air collisions between cooperating effectors.

To maintain clarity of exposition, we assume throughout this paper that a WTA layer is present and provides each effector with a fixed target assignment. Nevertheless, the proposed approach is directly applicable to dynamically changing assignments as well. At the engagement level, we adopt proportional navigation guidance (PNG) as the nominal law due to its simplicity, robustness, and widespread use in missile systems. PNG is a classical guidance law designed to steer a pursuer (e.g., missile or interceptor) towards a moving target by nullifying the line-of-sight (LOS) angular rate. However, the safety framework presented herein is not limited to PNG and can be coupled with any guidance law that generates acceleration commands.

Let the norm of the distance and the unit vector along the LOS between the target and the effector be defined as
\begin{align}
    r &= \|\mathbf{r}_{ik}\|, \\
    \hat{\mathbf{r}} &= \frac{\mathbf{r}_{ik}}{r}.
\end{align}
The LOS angular rate vector is given by
\begin{equation}
    \dot{\boldsymbol{\lambda}} = \frac{\mathbf{r}_{ik} \times \mathbf{v}_{ik}}{r^2}.
\end{equation}

Each effector \( i \) attempts to intercept an assigned target \( k \), using pure PNG. The pure PNG law prescribes a commanded acceleration perpendicular to the LOS direction, given by
\begin{equation}
    \mathbf{a}_{\text{nom},i} 
    = N \cdot \|\mathbf{v}_{ik}\| \cdot 
    \left( \dot{\boldsymbol{\lambda}} \times \hat{\mathbf{r}} \right),
\end{equation}
where \( N > 0 \) is the navigation constant and \( \|\mathbf{v}_{ik}\| \) is the relative closing speed. 
The acceleration vector lies orthogonal to the LOS and serves to nullify the LOS rate, aligning the velocity vector of the effector with the LOS over time.

While PNG provides efficient interception trajectories, its standard formulation does not explicitly consider interactions among multiple effectors. Consequently, agents pursuing different targets may inadvertently converge into close proximity, creating the risk of mutual collision. This limitation motivates the integration of a real-time safety filter that modifies the nominal guidance command into a safe acceleration command \( \mathbf{a}_i \), thereby ensuring both feasibility and satisfaction of safety constraints.

%The proposed safety filter is formulated as a decentralized Quadratic Program (QP), in which each agent considers both its local state and the states of nearby agents within a critical neighborhood \( r_{\text{neigh}} \). The objective is to remain as close as possible to the nominal command, while ensuring safety via barrier constraints.% To handle violations of these constraints due to modeling errors or physical saturation, we propose an extension that activates a CLF-based recovery condition when standard CBF constraints are unsatisfiable.

\section{Decentralized Robust CBF-based QP Formulation for Collision Avoidance}

To reduce the computational complexity in multi-agent collision avoidance, each agent \( i \in \mathcal{M} \) limits its safety considerations to a local neighborhood defined by
\[
\mathcal{N}_i := \left\{ j \in \mathcal{M} \setminus \{i\} \,\middle|\, \| \mathbf{p}_i - \mathbf{p}_j \| \leq r_{\text{neigh}} \right\},
\]
where \( r_{\text{neigh},i} > 0 \) denotes the maximum interaction radius within which safety constraints are actively considered. This localized filtering reduces the number of constraints each agent must evaluate, enabling real-time feasibility in large-scale systems, without compromising safety for a sufficiently large neighborhood $\mathcal{N}_i$ defined via \( r_{\text{neigh},i} \) \cite{BORRMANN2015UAV}.

To further reduce the burden of unnecessary constraint evaluation, the safety filter is activated for neighbor \( j \in \mathcal{N}_i \) only when two concurrent conditions are met: 
\begin{itemize}
    \item First, the neighbor must be within a predefined activation range \( r_{\text{crit}} \), such that: \( \| \mathbf{r}_{ij} \| \leq r_{\text{crit}} \), for \( r_{\text{crit}} \leq r_{\text{neigh}} \).
    \item Second, the predicted miss distance, referred to as the zero-effort miss (ZEM), must fall below a critical threshold: \( \text{ZEM}_{ij} \leq \eta r_{\text{crit}} \), with \(0 < \eta < 1\). 
\end{itemize}

To anticipate potential future collisions, we compute the ZEM distance, which quantifies the predicted minimum separation between two agents assuming they maintain their current velocities. This quantity is derived from the relative motion between agents \( i \) and \( j \), under the assumption of constant velocity propagation. The relative position at time \( t = t_0 + T \) evolves as
\begin{equation}
    \mathbf{r}_{ij}(t) = \mathbf{r}_{ij}(t_0) + T \mathbf{v}_{ij}(t_0).
\end{equation}

To compute the time \( T \) at which the minimum separation occurs, we minimize the squared distance
\begin{equation}
    D_{ij}^2(t) := \| \mathbf{r}_{ij}(t) \|^2 = \left\| \mathbf{r}_{ij}(t_0) + T \mathbf{v}_{ij}(t_0) \right\|^2.
\end{equation}
This expands to
\begin{equation}
    D_{ij}^2(t) = \| \mathbf{r}_{ij}(t_0) \|^2 + 2T \mathbf{r}_{ij}(t_0)^\top \mathbf{v}_{ij}(t_0) + T^2 \| \mathbf{v}_{ij}(t_0) \|^2.
\end{equation}

The time-to-go until closest approach, denoted \( T_{\text{ZEM}} \), is found by setting the derivative of \( D_{ij}^2(t) \) with respect to \( T \) to zero:
\begin{equation}
    \frac{d}{dT} D_{ij}^2(t) = 2 \mathbf{r}_{ij}(t_0)^\top \mathbf{v}_{ij}(t_0) + 2T \| \mathbf{v}_{ij}(t_0) \|^2 = 0.
\end{equation}
Solving yields:
\begin{equation}
    T_{\text{ZEM}} = -\frac{\mathbf{r}_{ij}(t_0)^\top \mathbf{v}_{ij}(t_0)}{\| \mathbf{v}_{ij}(t_0) \|^2}.
\end{equation}

If \( T_{\text{ZEM}} > 0 \), the agents are on a converging trajectory, and the point of closest approach lies in the future. If \( T_{\text{ZEM}} < 0 \), the agents are on a diverging path, and their minimum separation already occurred in the past; thus, a collision is not imminent under current motion assumptions.

Substituting \( T_{\text{ZEM}} \) back into the relative position yields the predicted miss distance:
\begin{equation}
    \text{ZEM}_{ij} := \left\| \mathbf{r}_{ij}(t_0) + T_{\text{ZEM}} \mathbf{v}_{ij}(t_0) \right\|.
\end{equation}

This value is used in the event-triggered safety filter activation logic. Specifically, if both \( T_{\text{ZEM}} > 0 \) and \( \text{ZEM}_{ij} \leq \eta r_{\text{crit}} \), with \( \eta \in (0,1) \) a design constant, the safety filter is activated to resolve the predicted future violation of the safety margin.

We define a CBF candidate to encode the safety constraint between each agent \( i \) and a neighbor \( j \). Let the minimum safety distance \( r_s \), defined such that \( \eta r_{\text{crit}} > r_s > 0 \), be given, and require that the following safety condition holds at all times
\begin{align}
\|\mathbf{r}_{ij}\|^2 \ge r_s^2.
\end{align}

In the following, the analysis for the derivation of a valid RCBF is conducted under the assumption of a one-on-one collision avoidance scenario. It should be emphasized, however, that this represents only a sub-approximation of the actual multi-agent problem, where an RCBF would, in principle, need to be derived for each possible interaction case. Nevertheless, it will be shown that, in most situations, distinct safety constraints for multiple agents within the relevant neighborhood, derived from the considered one-on-one collision avoidance scenario, can be dynamically incorporated as separate constraints into a safety filter, depending on the current interaction situation. Moreover, it will be later shown that, when such constraints are conflicting, a slack-variable-based prioritization of the most critical conflicts can still ensure feasibility of the overall safety problem, albeit at the expense of relaxing the theoretical guarantees for less critical safety constraints at the current time step.\\

\noindent We define the following barrier function as 
\begin{align} 
h_{ij}(\mathbf{x}) = -\|\mathbf{r}_{ij}\|^2 +r_s^2 \le 0 
\end{align} 
and the corresponding safe set as
\begin{align} 
{\mathcal{S}}_{ij}
=
\{
\mathbf{x} \in \mathbb{R}^n \mid 
	h_{ij}(\mathbf{x})\le 0
\}.
\end{align} 

To verify the existence of a valid RCBF, we first need to calculate $\ddot h_{ij}(\mathbf{x})$ to check the existence of {\(\xi>0\)} according to Theorem \ref{num:theorem_theo9_rcbf_paper}. Taking the time derivatives, it follows that 
\begin{align} 
\dot h_{ij}(\mathbf{x}) = -2\mathbf{r}_{ij}^\top \mathbf{v}_{ij}, 
\end{align} 
\begin{align} 
\ddot h_{ij}(\mathbf{x}) = -2\|\mathbf{v}_{ij}\|^2 -2\mathbf{r}_{ij}^\top(\mathbf{a}_i-\mathbf{a}_j). 
\label{eq:second_time_derivative_h}
\end{align} 
%\textcolor{blue}{Comment: Do you think this is the best way to define the input space? I was thinking about something like,  \(\mathbf{a}_i\in \mathcal{U}=\{\mathbf{a}\in\mathbb{R}^3 \mid \|\mathbf{a}\|\le a_{\max}\}\). } 
We assume the available control inputs are constrained within a sphere \(\mathbf{a}_i\in \mathcal{U}=\{\mathbf{a}\in\mathbb{R}^3 \mid \|\mathbf{a}\|\le {a_{\max}}\}\). To establish the worst-case bound according to (\ref{num:rcbf_max_acc}), we have to evaluate 
\begin{align} 
\inf_{\mathbf{a}_i,\mathbf{a}_j\in \mathcal{U}} \left( 
\ddot h_{ij}(\mathbf{x})\right)
=
\inf_{\mathbf{a}_i,\mathbf{a}_j\in \mathcal{U}} \left( -2\|\mathbf{v}_{ij}\|^2 -2\mathbf{r}_{ij}^\top\mathbf{a}_i +2\mathbf{r}_{ij}^\top\mathbf{a}_j \right) \nonumber
\end{align} 
for all $\mathbf{x}\in {\mathcal{S}}_{ij}$.
To minimize this expression, we need $\mathbf{r}_{ij}^\top\mathbf{a}_i$ to become maximal with positive sign, because then the term $ -2\mathbf{r}_{ij}^\top\mathbf{a}_i$ is minimal with respect to control input $\mathbf{a}_i$. The scalar product becomes maximal positive when $\mathbf{a}_i$ is aligned with $\mathbf{r}_{ij}$ and points in the same direction. This is intuitive, as $\mathbf{r}_{ij}$ points away from agent $j$ toward agent $i$, and so the acceleration of agent $i$ should ideally point in the same direction to maximize separation. This represents the optimal control input to avoid the constraint violation. However, the input must respect the physical actuation constraints. Therefore, we select $\mathbf{a}_i=\frac{\mathbf{r}_{ij}}{\|\mathbf{r}_{ij}\|}{a_{\max}}$, which is aligned with $\mathbf{r}_{ij}$ but scaled to the boundary of the input constraint sphere. Consequently, 
\begin{align}
&\inf_{\mathbf{a}_i,\in \mathcal{U}} \left( -2\|\mathbf{v}_{ij}\|^2 -2\mathbf{r}_{ij}^\top\mathbf{a}_i +2\mathbf{r}_{ij}^\top\mathbf{a}_j \right) \nonumber\\
&\quad = -2\|\mathbf{v}_{ij}\|^2 -2\|\mathbf{r}_{ij}\|{a_{\max}} +2\mathbf{r}_{ij}^\top\mathbf{a}_j. 
\end{align} 
Now we must make assumptions regarding the control input of agent $j$. From a cooperative collision avoidance perspective, it would be beneficial to minimize $2\mathbf{r}_{ij}^\top\mathbf{a}_j$, as this allows us to select a larger value of {\(\xi\)}  in (\ref{num:rcbf_max_acc}), thereby maximizing the subset of the safe set that can be rendered forward invariant. The minimum value of $2\mathbf{r}_{ij}^\top\mathbf{a}_j$ is achieved when $\mathbf{a}_j$ points opposite to $\mathbf{r}_{ij}$. This makes physical sense because in this case the acceleration of agent $j$ points away from agent $i$, contributing to increased separation. Respecting the input constraints of agent $j$, we have $\mathbf{a}_j=-\frac{\mathbf{r}_{ij}}{\|\mathbf{r}_{ij}\|}{a_{\max}}$. Consequently, 
\begin{align}
&\inf_{\mathbf{a}_i,\mathbf{a}_j\in \mathcal{U}} \left( -2\|\mathbf{v}_{ij}\|^2 -2\mathbf{r}_{ij}^\top\mathbf{a}_i +2\mathbf{r}_{ij}^\top\mathbf{a}_j \right) \nonumber\\
&\quad =-2\|\mathbf{v}_{ij}\|^2 -2\|\mathbf{r}_{ij}\|{a_{\max}}-2\mathbf{r}_{ij}^\top\frac{\mathbf{r}_{ij}}{\|\mathbf{r}_{ij}\|}{a_{\max}}\nonumber\\
&\quad = -2\|\mathbf{v}_{ij}\|^2 -4\|\mathbf{r}_{ij}\|{a_{\max}}. 
\end{align} Further, we obtain 
\begin{align}
-2\|\mathbf{v}_{ij}\|^2 -4\|\mathbf{r}_{ij}\|{a_{\max}} \le -4\|\mathbf{r}_{ij}\|{a_{\max}}, 
\end{align}
and for the relative distance $\|\mathbf{r}_{ij}\|$ we know from the definition of the safe set that it is at least $\|\mathbf{r}_{ij}\| \ge r_s$. Combining these results, we finally derive $\forall \mathbf{x}\in {\mathcal{S}}_{ij}\colon$
\begin{align}
&\inf_{\mathbf{a}_i,\mathbf{a}_j\in \mathcal{U}} \left( -2\mathbf{v}_{ij}^\top\mathbf{v}_{ij} -2\mathbf{r}_{ij}^\top(\mathbf{a}_i-\mathbf{a}_j) \right) \nonumber\\
&\quad \le 
-4\|\mathbf{r}_{ij}\|{a_{\max}}
\le 
-4r_s{a_{\max}} \stackrel{!}{\le} - {\xi} 
\end{align} 
and {\(\xi>0\)} .
Consequently, the feasibility of the RCBF can always be ensured and the admissible range of the tuning parameter {\(\xi\)} is 
\begin{align}
0<
{\xi}
\le
4r_s{a_{\max}}.
\end{align} 
Note, that by taking the maximum value of {\(\xi\)}, i.\,e. {\(\xi=
4r_sa_{\max}\)}, we can maximize the subset ${\mathcal{S}}_{ij}^{res}$ defined by
\begin{align}
{\mathcal{S}}_{ij}^{res}
=
\{
\mathbf{x} \in \mathbb{R}^n \mid 
	H_{ij}(\mathbf{x})\le 0 \land
    h_{ij}(\mathbf{x})\le 0
\}
\subseteq {\mathcal{S}}_{ij}
\end{align} 
that can be rendered invariant under input constraints.

Having derived the necessary feasibility conditions, we now formally construct the RCBF for the pairwise collision avoidance scenario. The RCBF is given by
\begin{align}
	H_{ij}(t,\mathbf{x}) &= H_{ij}=h_{ij}+\frac{|\dot h_{ij}|\dot h_{ij}}{2{\xi}},
\end{align}
and under the feasibility conditions derived beforehand, we can guarantee that for all $\mathbf{x}\in {\mathcal{S}}_{ij}^{res}$ 
and any extended class-$\mathcal{K}$ function $\alpha\in\mathcal{K}$ the RCBF condition
\begin{align}
	\dot{H}_{ij}(t,\mathbf{x},\mathbf{a}_i,\mathbf{a}_j) 
      \;\leq\; \alpha\!\left(-H_{ij}(t,\mathbf{x})\right)
\end{align}
is always feasible under the input constraints. As a consequence, the subset ${\mathcal{S}}_{ij}^{res}$ can be rendered invariant under input constraints.

For the 1-by-1 collision avoidance problem under consideration, the RCBF condition can be explicitly stated as
\begin{align}
		\dot{H}_{ij}(t,\mathbf{x},\mathbf{a}_i,\mathbf{a}_j) 
        = &\dot{h}_{ij}(t,\mathbf{x}) \notag\\
        &+ \frac{|\dot{h}_{ij}(t,\mathbf{x})|}{{\xi}} \ddot{h}_{ij}(t,\mathbf{x},\mathbf{a}_i,\mathbf{a}_j),
\end{align}
With the second-order derivative given by Eq.~\eqref{eq:second_time_derivative_h}, which—upon substitution and rearrangement—leads to
\begin{align}
  -2\lvert \dot h\rvert\,\mathbf{r}_{ij}^\top \mathbf{a}_i
  \;&\le\;
  {\xi}\bigl(\alpha(-H)-\dot h\bigr)\notag\\
  & \qquad \quad
  + 2\lvert \dot h\rvert\,\|\mathbf{v}_{ij}\|^{2}
  - 2\lvert \dot h\rvert\,\mathbf{r}_{ij}^\top \mathbf{a}_j .
\end{align}
Under the cooperative collision avoidance assumption, we have $\mathbf{a}_j=-\mathbf{a}_i$, yielding
\begin{align}
		-4|\dot h|\mathbf{r}_{ij}^\top\mathbf{a}_i
		\le
        {\xi}\bigl(\alpha(-H)-\dot h\bigr)
        +2|\dot h|\|\mathbf{v}_{ij}\|^2.
\end{align}

This represents the most favorable case, since both agents actively accelerate away from each other, thereby maximizing the braking effect in the barrier condition.  In our implementation, this assumption is being evaluated online. Specifically, when a valid communication link to a neighboring effector is available, we assume cooperativeness and set \(\mathbf{a}_j=-\mathbf{a}_i\), which corresponds to the optimal avoidance behavior. If no valid connection exists, we instead assume \(\mathbf{a}_j=\mathbf{0}\), reflecting that the neighbor maintains its current motion without adversarial intent. This compromise avoids overly conservative modeling, such as presuming worst-case adversarial maneuvers, while still preserving safety guarantees consistent with the interception mission objectives.

Finally, the safety filter is implemented as a QP that modifies the nominal acceleration command \( \mathbf{a}_{\text{nom},i}(t) \). At each time step, the corrected safe input \( \mathbf{a}_i(t) \) is obtained by solving
\begin{align}
   &\min_{\mathbf{a}_i \in \mathbb{R}^3}
      \;\; \lVert \mathbf{a}_i - \mathbf{a}_{\text{nom},i} \rVert_W^2 \notag \\
   &\text{s.t.} \notag \\
   & \dot{H}_{ij}(t,\mathbf{x},\mathbf{a}_i,\mathbf{a}_j) 
      \;\leq\; \alpha\!\left(-H_{ij}(t,\mathbf{x})\right),
      \quad \forall j \in \mathcal{N}_i, 
      \label{eq:rcbf_constraint}
\end{align}
with \( \lVert \cdot \rVert_W^2 := (\cdot)^\top W (\cdot) \) denoting a weighted norm defined by a symmetric positive definite matrix \( W \in \mathbb{R}^{3\times3} \). Constraint \eqref{eq:rcbf_constraint} enforces forward invariance of the robust control barrier function for every active neighbor \( j \in \mathcal{N}_i \). Since the derived RCBFs inherently encode input constraints in the one-on-one case, feasibility of the QP is guaranteed under the conditions established in the previous subsection, without requiring additional explicit acceleration bounds.

As noted previously, in multi-agent scenarios the distinct safety constraints for all agents within the relevant neighborhood, derived from the one-on-one collision avoidance formulation, can be dynamically incorporated as separate constraints into the safety filter, depending on the current interaction situation. This may, however, result in conflicting constraints. To address such cases, the next section introduces a slack-variable-based prioritization scheme that ensures feasibility of the overall safety problem, albeit at the expense of relaxing the theoretical guarantees for less critical safety constraints at the current time step.

\section{Criticality-Based Prioritization under Conflicting Constraints}
For the proposed decentralized, event-triggered safety filter framework, each agent autonomously evaluates and activates pairwise barrier-function constraints depending on the local interaction state. While individual pairwise constraints rigorously certify safety in isolation, their simultaneous enforcement does not automatically extend to the multi-agent setting. In particular, situations may arise where multiple constraints are active at once and form mutually restrictive conditions. A representative example is the so-called leaky-corner problem, in which the intersection of several pairwise safe sets leaves no admissible input, even though the system is physically capable of maintaining safe separation. More broadly, the lack of compositionality between local certificates and the global safe set can induce temporary deadlocks that undermine feasibility without necessarily reflecting true collision risk \cite{Aloor_2025}.

Constraint aggregation offers one mitigation strategy by merging multiple neighbors into virtual obstacles, thereby reducing the number of active constraints and the QP dimension. While effective in limiting conflicts, this approach introduces conservatism because aggregated safe sets are typically inflated, which can waste maneuvering authority and even cause infeasibility if the required evasive action exceeds available acceleration limits. Furthermore, online aggregation must ensure that the agent’s own state is not inadvertently enclosed, which adds geometric complexity.

A more flexible alternative is slack-variable relaxation, which retains all pairwise constraints but allows principled violations through slack variables:
\begin{equation}
\label{eq:qp-slack}
\begin{aligned}
& \min_{\mathbf{a}_i\in \mathbb{R}^3,\{\delta_{ij}\ge 0\}} \quad 
    \|\mathbf{a}_i - \mathbf{a}_{\text{nom},i}\|_W^2 
    + \sum_{j \in \mathcal{N}_i} w_{ij}\,\delta_{ij} \\
&\text{s.t.} \quad 
\dot{H}_{ij}(t,\mathbf{x},\mathbf{a}_i,\mathbf{a}_j) 
    \le \alpha(-H_{ij}(t,\mathbf{x})) + \delta_{ij}, 
    \quad \forall j \in \mathcal{N}_i.
\end{aligned}
\end{equation}

Here, the auxiliary variables $\delta_{ij}\ge 0$ represent constraint relaxations, and the weights $w_{ij}>0$ encode interaction priorities. Large weights make violations costly, biasing the optimizer to enforce critical constraints strictly, while smaller weights allow limited relaxation on less critical ones. From a multi-objective perspective, it could be argued that the solution is Pareto optimal, in the sense that the approach seeks to avoid constraint violations whenever possible \cite{Cao_2025}. When violations are unavoidable, no single violation can be reduced without either worsening another or deviating further from the nominal guidance command. In such cases, residual violations are applied preferentially to less critical constraints. In the considered case, priorities are assigned in a state-dependent manner:
\[
w_{ij} = w_0 
+ \frac{k_d}{\varepsilon+\|\mathbf{r}_{ij}\|} 
+ \frac{k_t}{\varepsilon+T^+_{\mathrm{ZEM},ij}},
\]
where $T^+_{\mathrm{ZEM},ij}=\max\{0,T_{\mathrm{ZEM},ij}\}$. The distance-dependent term increases the weight for nearby neighbors, while the time-to-go term emphasizes encounters that are closing rapidly, thereby incorporating relative velocity information. The  tuning parameters $k_d>0$ and $k_t>0$ determine their relative influence. The constant baseline $w_0$ ensures a minimum prioritization level, and $\varepsilon>0$ prevents singularities. In practice, this means that close or imminent constraints are unlikely to be relaxed, while distant or non-threatening ones absorb slack when conflicts occur. For example, if two constraints are in conflict, the optimization will strictly enforce the one corresponding to a rapidly approaching neighbor while relaxing the one associated with a far-away agent.

%This slack-based relaxation enlarges the feasible set while keeping control inputs within actuator limits, and concentrates residual violations on less critical neighbors. Complete feasibility cannot be guaranteed under severe saturation when control authority is insufficient, but this risk is mitigated through event-triggered activation (which limits the number of simultaneously active constraints), robust RCBF formulations that account for worst-case neighbor behaviors, and appropriate tuning of the safety distance $r_s$ relative to velocity and acceleration bounds.
%\section{HOCBF-Based Safety Filtering with Neighborhood Triggering for Cooperative Interceptors}
%\input{sections/Main_Contribution}
\section{Simulation Results}
To evaluate the proposed decentralized RCBF-based safety filter, we consider a three-on-three interception scenario with PNG serving as the nominal law. Each effector is assigned to a unique target, resulting in one-to-one pairings throughout the engagement. The evaluation compares two cases: (i) a baseline scenario using PNG-only guidance and (ii) a scenario where PNG is combined with the proposed safety filter. 

The initial conditions for all agents are summarized in Table~\ref{tab:init_conditions}. The effectors are launched from distinct lateral positions with forward velocities around $300\,\text{m/s}$. Each effector is subject to a maximum speed of $v_{\max}=306\,\text{m/s}$ and a maximum lateral acceleration of $a_{\max}=40g$. The targets are positioned approximately $25\,$km ahead of the effectors and move with constant velocities, i.e., they are non-accelerating. In all cases, the vertical velocity components are initialized as zero, confining the engagement to the horizontal plane. 

\begin{table}[h]
\centering
\caption{Initial states of effectors and targets.}
\label{tab:init_conditions}
\begin{tabular}{c c c}
\toprule
Agent & $\mathbf{p}(t_0) = [x,y,z]^\top$ [m] & $\mathbf{v}(t_0) = [v_x,v_y,v_z]^\top$ [m/s] \\
\midrule
Missile 1 & $[0,\,-100,\,0]^\top$   & $[300,\,10,\,0]^\top$ \\
Missile 2 & $[0,\,100,\,0]^\top$    & $[300,\,-10,\,0]^\top$ \\
Missile 3 & $[0,\,-500,\,0]^\top$   & $[300,\,60,\,0]^\top$ \\
\midrule
Target 1   & $[25600,\,-320,\,0]^\top$   & $[-320,\,64,\,0]^\top$ \\
Target 2   & $[25600,\,320,\,0]^\top$    & $[-640,\,0,\,0]^\top$ \\
Target 3   & $[25600,\,-1600,\,0]^\top$  & $[-640,\,-32,\,0]^\top$ \\
\bottomrule
\end{tabular}
\end{table}
\begin{figure}
     \centering
     \includegraphics[width=\linewidth]{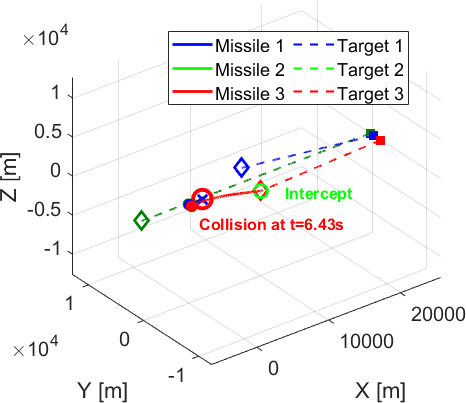}
     \caption{3D trajectories with PNG only. Missiles~1 and~2 collide at $t=6.43\,$s.}
     \label{fig:3d_no_filter}
\end{figure}

\begin{figure}
     \centering
     \includegraphics[width=\linewidth]{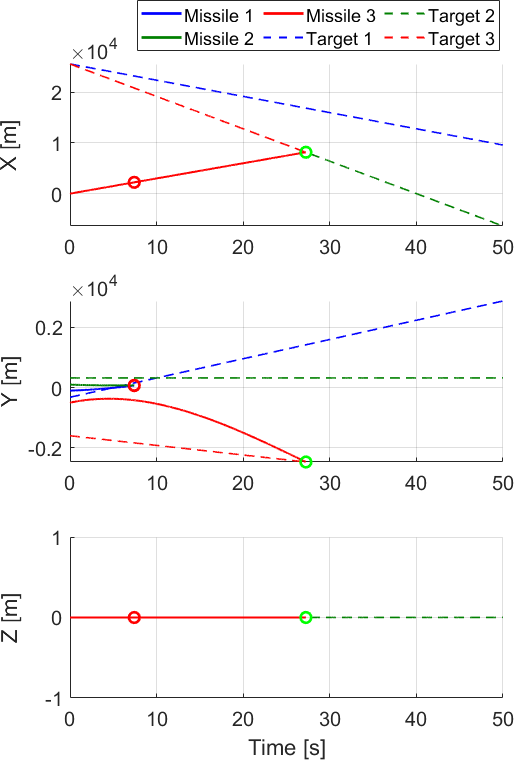}
     \caption{Position histories with PNG only. Collision occurs between Missiles~1 and~2.}
     \label{fig:pos_no_filter}
\end{figure}

\begin{figure}
     \centering
     \includegraphics[width=\linewidth]{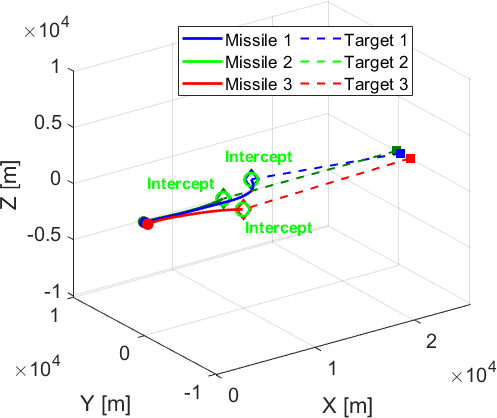}
     \caption{3D trajectories with PNG + safety filter. All missiles intercept safely.}
     \label{fig:3d_rcbf}
\end{figure}

\begin{figure}
     \centering
     \includegraphics[width=\linewidth]{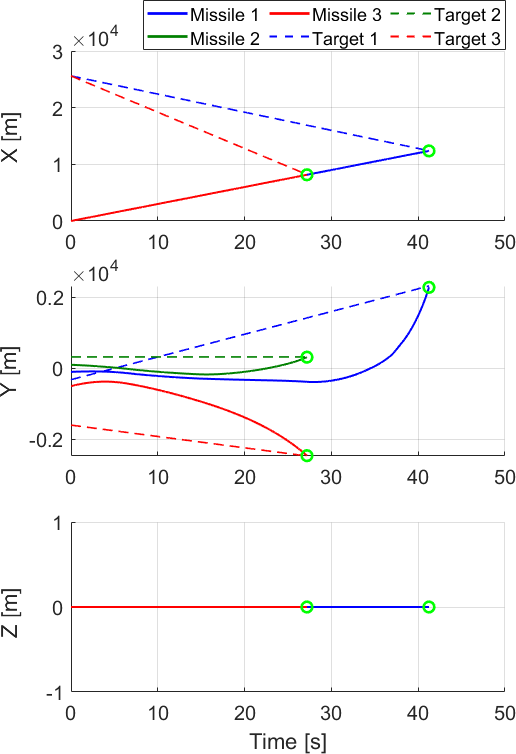}
     \caption{Position histories with PNG + safety filter. Separation maintained.}
     \label{fig:pos_rcbf}
\end{figure}

Figures~\ref{fig:3d_no_filter} and~\ref{fig:pos_no_filter} illustrate the outcome of the baseline PNG-only case. While each missile pursues its assigned target, Missiles~1 and~2 approach dangerously close and collide at $t=6.43\,$s. This occurs because PNG does not account for inter-agent interactions, and although Missile~3 intercepts its target successfully, the overall mission is compromised by the mid-air collision of the other two effectors.  

In contrast, Figures~\ref{fig:3d_rcbf} and~\ref{fig:pos_rcbf} show the case with the proposed safety filter. Here, the decentralized RCBF constraints are dynamically enforced for relevant neighbors, and as Missiles~1 and~2 approach, the safety filter modifies their acceleration commands just enough to maintain separation while still guiding them toward interception. Both effectors achieve successful target engagements without collision. Importantly, Missile~3, which is well separated from the potential conflict, follows nearly the same trajectory as in the PNG-only case, demonstrating that the safety filter does not impose unnecessary conservatism on uninvolved agents.  

The comparison clearly shows that without the safety filter collisions can occur, even when target assignments are distinct, whereas with the filter all effectors remain safe and complete their assigned interceptions. Furthermore, the corrective action of the safety filter is localized and minimal, intervening only when conflicts arise, and preserving the nominal performance for effectors not directly involved in potential collisions. This ensures that the proposed approach provides both safety and efficiency, making it suitable for large-scale cooperative aerospace scenarios.

%\section{INTERGRATION SECTION}
%\input{sections/Integration}
\section{Conclusion}
We proposed a decentralized, event-triggered safety filter for cooperative guidance based on robust control barrier functions with explicit input constraints. A slack-variable prioritization scheme ensures feasibility under conflicting constraints, while range- and ZEM-based triggering enables scalability. Simulations in many-on-many interception scenarios confirm that the method maintains collision-free operation and preserves nominal guidance performance, providing an efficient and scalable solution for safety-critical multi-agent aerospace systems.

\bibliographystyle{IEEEtran}
\bibliography{./references} 

\end{document}